\documentclass[10pt,conference,letterpaper]{IEEEtran}
\IEEEoverridecommandlockouts
\usepackage[utf8]{inputenc}
\usepackage[english]{babel}
\usepackage{multirow} 
\usepackage{booktabs}
\usepackage{threeparttable} 
\usepackage{cite}
\usepackage{amsmath,amssymb,amsfonts}
\usepackage{algorithmic}
\usepackage{graphicx}
\usepackage{textcomp}
\usepackage{xcolor}
\def\BibTeX{{\rm B\kern-.05em{\sc i\kern-.025em b}\kern-.08em
    T\kern-.1667em\lower.7ex\hbox{E}\kern-.125emX}}

\usepackage{balance}
\usepackage[ruled, lined, linesnumbered, commentsnumbered, noend]{algorithm2e}

\usepackage{tikz}
\usetikzlibrary{positioning,shapes.geometric}
\usetikzlibrary{positioning,intersections}

\usepackage{subcaption}
\usepackage{newfloat}
\DeclareFloatingEnvironment[fileext=frm,placement={tbp},name=Problem]{problem}
\graphicspath{{./grafiken/}}

\usepackage[nolist]{acronym}
\usepackage{hyperref}
\hypersetup{nolinks=true} 

\usepackage[absolute]{textpos} 
\tikzset{
  neuron/.style={
    circle ,
    inner sep=0pt,
    minimum width=5mm,
    draw=black
  }
}

\tikzset{
  empty/.style={
    circle ,
    inner sep=0pt,
    minimum width=5mm,
    draw=white
  }
}
\definecolor{nyellow}{RGB}{255,185,15}
\definecolor{ngreen}{RGB}{0,139,0}

\newsavebox{\jamBox}
\newlength{\jamWidth}
\newcommand{\jamIfToBig}[2]{%

    \savebox{\jamBox}{#2}%
    \settowidth{\jamWidth}{\usebox{\jamBox}}%

    \ifthenelse{\jamWidth < #1}%
        {\usebox{\jamBox}}%
        {\resizebox{#1}{!}{\usebox{\jamBox}}%
    }%
}

\begin{document}

\title{Green Segment Routing for Improved Sustainability of Backbone Networks}

\def\IEEEauthorrefmark#1{\raisebox{0pt}[0pt][0pt]{\textsuperscript{\footnotesize\ensuremath{\ifcase#1\or *\or \dagger\or \ddagger\or%
    \bullet\or \circ\or \cdot\or \times\or \checkmark\or **\or \dagger\dagger%
    \or \ddagger\ddagger \else\textsuperscript{\expandafter\romannumeral#1}\fi}}}}

\author{
  \IEEEauthorblockN{Daniel Otten\IEEEauthorrefmark{5}, Alexander Brundiers\IEEEauthorrefmark{5}, Timmy Sch\"{u}ller\IEEEauthorrefmark{4}, Nils Aschenbruck\IEEEauthorrefmark{5}}
\vspace*{.18cm}
\IEEEauthorblockA{
  \begin{tabular}{ccc}
    {\IEEEauthorrefmark{5}Osnabr\"{u}ck University, Institute of Computer Science} && 		   {\IEEEauthorrefmark{4}Deutsche Telekom Technik GmbH}\\
    {Friedrich-Janssen-Str. 1, 49076 Osnabr\"{u}ck, Germany} && {Gartenstraße 217, 48147 M\"{u}nster, Germany}\\
    {Email: \{daotten, aschenbruck,brundiers\}@uos.de} && {Email: timmy.schueller@telekom.de}\\
  \end{tabular}
}}

\maketitle

\acrodefplural{PoP}[PoPs]{Points of Presence}
\begin{acronym}
\acro{DEFO}{Declarative and Expressive Forwarding Optimizer}
\acro{ECMP}{Equal Cost Multipath}
\acro{IGP}{Interior Gateway Protocol}
\acro{IE}{Ingress-Egress}
\acro{ISP}{Internet Service Provider}
\acro{IS-IS}{Intermediate System to Intermediate System}
\acro{LDP}{Label Distribution Protocol}
\acro{LER}{Label Edge Router}
\acro{LP}{Linear Program}
\acro{LSR}{Label Switched Router}
\acro{LSP}{Label Switched Path}
\acro{MCF}{Multicommodity Flow}
\acro{MLU}{Maximum Link Utilization}
\acro{MO}{Midpoint Optimization}
\acro{MPLS}{Multiprotocol Label Switching}
\acro{MSD}{Maximum Segment Depth}
\acro{NDA}{non-disclosure agreement}
\acro{OSPF}{Open Shortest Path First}
\acro{PoP}{Point of Presence}
\acro{RSVP}{Resource Reservation Protocol}
\acro{RL2TLE}{Router-Level 2TLE}
\acro{SDN}{Software-defined Networking }
\acro{SC2SR}{Shortcut 2SR}
\acro{SID}{Segment Identifier}
\acro{SPR}{Shortest Path Routing}
\acro{SR}{Segment Routing}
\acro{SRLS}{Segment Routing Local Search}
\acro{TE}{Traffic Engineering}
\acro{TLE}{Tunnel Limit Extension}
\acro{WAE}{WAN Automation Engine}
\acro{w2TLE}{Weighted 2TLE}
\acro{ILP}{Integer Linear Program}
\end{acronym}

\begin{abstract}
Improving the energy efficiency of \ac{ISP} backbone networks is an important objective for \ac{ISP} operators.
In these networks, the overall traffic load throughout the day can vary drastically, resulting in many backbone networks being highly overprovisioned during periods of lower traffic volume.
In this paper, we propose a new \ac{SR}-based optimization algorithm that aims at reducing the energy consumption of networks during such low-traffic periods.
It uses the traffic steering capabilities of \ac{SR} to remove traffic from as many links as possible to allow the respective hardware components to be switched off.
Furthermore, it simultaneously ensures that solutions comply to additional operator requirements regarding the overall \acl{MLU} in the network.
Based on data from a Tier-1 \ac{ISP} and a public available dataset, we show that our approach allows for up to 70 \% of the overall linecards to be switched off, corresponding to an around 56 \% reduction of the overall energy consumption of the network in times of low traffic demands.

\end{abstract}

\acresetall

\section{Introduction}
In times of global warming and rising energy costs, reducing the energy consumption in various aspects of our lives has become a major objective across various branches of research.
Even though the Internet plays a crucial role in our everyday lives, it is often overlooked when it comes to identifying candidates that offer an energy saving potential.
\par
Especially \ac{ISP} networks which basically are the \textit{backbone} of the Internet offer quite some room for improvement when it comes to energy efficiency.
In order to ensure reliable service even under severe traffic load and/or failures, these networks are often highly over-provisioned in terms of capacity.
As a result, links are rarely utilized to their full capacity.
Especially during times with lower traffic volume (e.g., at night), the average link utilization can drop to 10 \% or lower \cite{isp_network_utilization}.
While such a design is a straightforward approach to ensure network reliability, it is not very efficient in terms of energy consumption and offers quite some room for improvement.
\par
With the difference between the maximum and minimum traffic volume per day being expected to increase to a factor of six \cite{ciscoreport}, shutting down parts of the hardware during these low-traffic periods is a promising approach for reducing the overall energy consumption in \ac{ISP} networks.
During such time periods, a substantial part of the network and the underlying hardware is actually not needed for operation and could be switched off in order to save energy.
However, in order to be able to seamlessly switch off network components without impacting the network's performance, traffic first needs to be routed away from those components.
The required traffic steering can be realized with various \ac{TE} approaches.
Over the recent years, multiple approaches have been proposed that leverage this idea of using \ac{TE} to reduce the energy consumption in backbone networks (e.g., \cite{Reducing-Power-Consumption, Zhang, green_sr}).
However, many of these approaches tend to have specific shortcomings that potentially prevent an effective practical deployment.
For example, some of them continuously switch components on and off to dynamically react to traffic changes.
While this works well in theory, operators are often hesitant to continuously alter or reconfigure their networks during normal operation. Instead, they prefer to have a stable configuration that is altered only a couple of times a day, at most.
Furthermore, many approaches rely on older \ac{TE} technologies like \ac{MPLS} with RSVP-TE \cite{rfc3209}, which induce unnecessarily high overhead into the network. Hence, network operators are switching to newer, light-weight technologies like \ac{SR} \cite{srArch}. 
\par
To address these issues, we propose a new \ac{SR}-based optimization algorithm for increasing the energy efficiency of backbone networks.
For a given network and traffic matrix, it finds \ac{SR} configurations that maximize the number of components that can be switched off while ensuring that a certain utilization threshold is not surpassed in the network.
These configurations are static and can be used for multiple hours during the daily low-traffic period without any dynamic adaptions.
Based on real-world data from a globally operating Tier-1 \ac{ISP}, we show that with the configurations computed by our algorithm, around 70 \% of linecards can be switched off for up to eight hours a day while still obtaining \ac{MLU} values of less than 70 \%.
This roughly translates to  50 \% reduction of the overall energy consumption in the network. 
\section{Background}\label{sec:Background}
This section introduces background information regarding the main aspects of this work: Energy consumption of \ac{ISP} backbone networks and \ac{SR}.
\subsection{Power Consumption of Backbone Networks} \label{subsec:power}
To be able to develop new approaches to reduce the power consumption of backbone networks, we need to understand how this consumption is made up.
Generally, the overall power consumption of a network equals to the sum of the power consumption of its hardware components.
Thereby, the large majority of energy is taken up by routers and switches.
Hence, a straightforward approach to reduce the energy consumption could aim at minimizing the number of active routers in the network, switching off idle ones.
However, in the context of backbone networks, switching off entire routers is often not feasible.
During low-traffic periods, the volume of traffic flows is often reduced, but rarely flows completely dry up. Hence, there is always at least some residual capacity required to route these flows or to account for slight traffic variations.
Furthermore, depending on the network topology, it might break down the network into multiple disconnected parts, which would be detrimental for network performance.
Particularly, at the network edge, routers typically connect the \ac{ISP} backbone to smaller customer networks and peering partners.
Switching these routers off would disconnect them from the \ac{ISP} network. Furthermore, booting a whole router can take up to $30$ minutes. This makes it impossible to react quickly when additional capacity is needed.
\par
Since shutting down entire routers is not feasible, the power consumption of individual routers should be minimized instead.
It comprises two main components: The power consumption of the chassis (e.g., cooling etc.) and the power consumption of the \textit{linecards}.
Linecards provide the physical endpoints for the connections between routers, the so-called ports.
Depending on the specific linecard model, the number of ports, as well as their capacity, can vary.
Modern linecards often feature between eight and twelve ports with a bandwidth of up to 100 Gbps per port.
Since linecards make up to 80 \% of the overall power consumption of a router \cite{power_consumption, cisco-power}, switching off idle linecards can result in a substantial reduction of the overall power consumption of a router and, hence, the whole network.
Many modern linecards (e.g., the Cisco ASR 9900) even come with a build in power-saving mode, which allows switching them on and off in an efficient and fast manner.\subsection{Segment Routing} \label{subsec:2sr}
\acl{SR} \cite{srArch} is a recent network tunneling technology based on the source-routing paradigm.
It allows specifying a list of waypoints (so-called segments) that a packet must visit in the given order.
Depending on the nature of the related waypoints, different types of segments can be used.
\textit{Node-segments}, for example, refer to a specific node (router) in the network, while \textit{adjacency-segments} identify links and \textit{service-segments} can be used to steer traffic into certain services (i.e., packet filters).
Despite the wide range of segment types, most \ac{SR}-based \ac{TE} approaches rely entirely on node-segments, as this simplifies optimization as well as practical operation while still offering traffic steering capabilities.
\par
In accordance to the source-routing paradigm, the respective segment list for a packet is added at the ingress node where it enters the \ac{SR} domain.
This way, a packet's path through the network is already predetermined at its time of entrance.
The forwarding paths towards intermediate segments are determined by the respective \ac{IGP} of the network.
When only considering node-segments, a \ac{SR} path basically is a concatenation of the multiple shortest paths between its segments.
While, in theory, arbitrarily many segments can be applied to a packet, this number is often limited by hardware constraints \cite{schueller_ton}.
Furthermore, each added segment adds a bit of additional overhead to the packet.
Hence, it is generally preferable to keep the number of segments as low as possible.
Depending on the maximum number $k$ of segments per path, this is called $k$-\ac{SR}.
While restricting the number of segments can, in theory, be a severe limitation of the traffic steering capabilities \cite{schueller_ton}, various research has shown that virtually optimal results can be obtained with just two or three segments (cf. \cite{schueller_ton} or \cite{defo2}).
\par
A major advantage of \ac{SR} compared to other \ac{TE} technologies is its significantly lower overhead.
For example, \ac{MPLS} tunnels that are realized with \ac{RSVP}-TE \cite{rfc3209} have to be configured and maintained on every node of the tunnel. This results in significant network overhead, especially if the number of tunnels in the network increases.
As a result, this approach does not scale well with network size.
Contrary to this, \ac{SR} is a stateless protocol that only requires configuration on the headend node of an \ac{SR} path, but not along the way. All other required information is encoded in the packet itself.
The only technical requirement for \ac{SR} is a special extension for the underlying \ac{IGP}. Other protocols, like the \ac{LDP} for \ac{MPLS}, are not required.
This substantially reduces the overhead that is introduced in the networks and makes \ac{SR} scale significantly better with network size.
The individual, per-flow traffic control paired with an exceptionally low overhead render \ac{SR} a premier tool for \ac{TE} purposes.
\section{Related Work} \label{sec:rel_work}

Around 20 years ago, it has been observed that the Internet requires a substantial amount of energy and that this demand is going to increase even further in the future.
With increasing energy costs and global warming becoming one of the prevalent problems of our time, more and more research is conducted with the goal of reducing the power consumption of the Internet. 

Many approaches like \cite{Energy_Aware_Managment}, \cite{Energy_Aware_Managment-SPR}, or \cite{Reducing-Power-Consumption} are not applicable in backbone networks, since
they assume that entire routers can be turned off. This, however, is not really feasible in practice. As already explained in Section II-A, shutting down edge-routers would result in a disconnection of customers. Furthermore, ISP operators are also quite hesitant when it comes to shutting down core-routers as this can have negative impact on the overall connectivity in the backbone, which becomes especially problematic in the face of failures.
\par
Other publications (e.g.,  \cite{Energy_Aware_Managment-SPR, Augmenting-EEE}) use \ac{IGP} metric tuning for traffic steering. The goal of metric tuning is to optimize the metrics within a network in order to reach a predefined target. Thus, it simply relies on the \ac{OSPF} \ac{IGP}.
It is still used for strategic \ac{TE} and long-term optimization, but short-term changes (i.e., routing traffic away from a link that has to be switched off for a few hours) are realized with other technologies like \ac{MPLS} or, more recently, \ac{SR}.
The reason for this is that metric tuning can have a lot of unwanted and negative side effects on the network.
Therefore, it should be done with care and as rarely as possible.
As a result, Green TE approaches that rely on metric tuning are not applicable in most modern networks.
\par
One of the most cited works in the context of Green \ac{TE} is \cite{Zhang}.
It presents an ILP-based algorithm that tries to minimize the power-consumption of a network by rerouting traffic to allow entire links to be switched off.
It is shown that this approach can reduce the power consumption by up to 42 \%.
However, it relies on \ac{MPLS} for traffic steering, which can result in significant network overhead for larger networks (cf. Section \ref{subsec:2sr}).
Modern approaches should utilize, \ac{SR} instead of \ac{MPLS} to benefit from its improved scalability.
The approach of \cite{Zhang} is not readily applicable towards a use with, \ac{SR} since \ac{MPLS} can be used to produce virtually arbitrary forwarding paths.
In contrast, \ac{SR}, when used with only a limited number of segments, is more limited in its traffic steering capabilities.
Furthermore, the work of \cite{Zhang} does not consider traffic changes over the course of a day. 
Evaluation is done on a network with a virtually constant traffic load throughout the day, which is an unrealistic assumption for modern \ac{ISP} networks (c.f. \cite{schueller_ton}).
\par
In addition, even for networks with less than 20 nodes, the authors already had to limit the computation.
For reference, modern \ac{ISP} backbone networks often comprise hundreds if not thousands of nodes.
Hence, we believe that the algorithm of \cite{Zhang} cannot be efficiently used for such large networks.
\par
A more recent approach to Green-\ac{TE} that utilizes \ac{SR}-based \ac{SDN} is \cite{green_sr}.
It is based on a simple algorithm that removes the least used link from the topology, while making sure that the network is not broken up into multiple parts. Moreover, it is ensured that a previously defined upper bound for the \ac{MLU} is not exceeded.
Before the removal of a link, a centralized \ac{SDN} controller computes the respective \ac{SR} configurations to free this link from traffic.
All of this is done in an automated and dynamic fashion, in which the \ac{SDN} controller continuously monitors the network and switches off links when possible, but also switches them on again if a certain \ac{MLU} threshold is surpassed.
It is shown that this approach can switch off up to 44 \% of links in a network.
\par
This approach was extended by  \cite{Per_Packet_based} to comply with the requirements of a Datacenter. Thus, this approach is optimized for so-called fat tree topologies. This topology usually does not occur in backbone networks. Another work, that combines \ac{SDN} and \ac{SR} for green-\ac{TE} is \cite{Intelligent_Path}. The focus of this paper is to minimize the number of SDN switches in the network while ensuring that the majority of flows can be controlled. 
Since this approach also relies on switching off whole routers, it is most likely not suitable for a use in an ISP backbone.
\par
Other approaches like \cite{GreyWolf} or  \cite{An_SDN_energy_saving_method} rely on a SDN controller too. They do not use \ac{SR} to steer traffic through the network. Instead, they allow arbitrary paths through the network. This can cause a lot of overhead, as we discussed before.
However, it strongly relies on the presence of a centralized controller that continuously monitors and alters the network and switches links on and off.
Often network operators are hesitant to deploy such automated and highly dynamic network re-configurations as this always features the risk of an accidental misconfiguration of the network.
This can have a detrimental effect on the overall network performance.
Instead, operators prefer solutions with a single stable configuration that can be preemptively checked for correctness and applicability by a human expert before rolling it out into the network.

\par
In this paper, we address the need for such approaches by proposing an algorithm that computes a single, stable \ac{SR} configuration that, once it is brought out into the network, can be used over multiple hours during low-traffic periods.

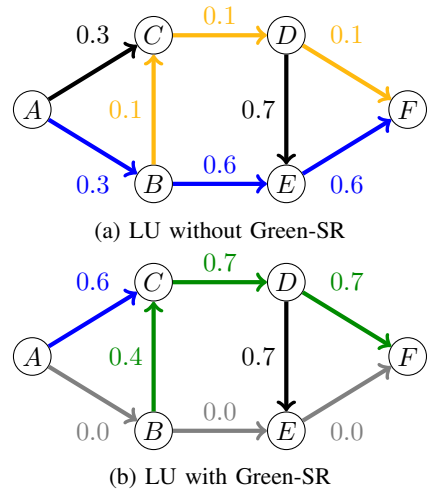
\begin{figure}
	\centering
	\begin{subfigure}{0.98\linewidth}
		\centering
   \begin{tikzpicture}
   \node[neuron,align=center] 
  (in1) {$A$};
  \node[neuron, below right = 0.625cm and 1.25 cm  of in1, align=center ] 
  (in2) {$B$};
    \node[neuron, above right = 0.625cm  and 1.25 cm  of in1, align=center ] 
  (in3) {$C$};
    \node[neuron, right = 1.25cm  of in3, align=center ] 
  (in4) {$D$};
      \node[neuron, right = 1.25cm  of in2, align=center ] 
  (in5) {$E$};
    \node[neuron, above right = 0.625cm and 1.25cm  of in5, align=center ] 
  (in6) {$F$};
  
 \draw[->, ultra thick,blue] (in1) to node[midway,below = 0.25cm]{$0.3$} (in2) ;
  \draw[->, ultra thick,black] (in1) to node[midway,above = 0.25cm]{$0.3$}(in3);
  
  \draw[->, ultra thick,blue] (in2) to node[midway,above]{$0.6$}(in5);
  \draw[->, ultra thick,nyellow] (in2) to node[midway,left]{$0.1$} (in3);
  
  \draw[->, ultra thick,nyellow] (in3) to node[midway,above]{$0.1$}  (in4);
  
  \draw[->, ultra thick,black] (in4) to node[midway,left]{$0.7$}  (in5) ;
  
  \draw[<-, ultra thick,nyellow] (in6) to node[midway,above = 0.25cm]{$0.1$} (in4);
  \draw[<-, ultra thick,blue] (in6) to node[midway,below = 0.25cm]{$0.6$} (in5);

\end{tikzpicture}
	\caption{LU without Green-SR}
	\label{subfig:ohne TE}
	\end{subfigure} \\

\begin{subfigure}{0.98\linewidth}
\centering
  \begin{tikzpicture}
     \node[neuron,align=center] 
  (in1) {$A$};
  \node[neuron, below right = 0.625cm and 1.25 cm  of in1, align=center ] 
  (in2) {$B$};
    \node[neuron, above right = 0.625cm  and 1.25 cm  of in1, align=center ] 
  (in3) {$C$};
    \node[neuron, right = 1.25cm  of in3, align=center ] 
  (in4) {$D$};
      \node[neuron, right = 1.25cm  of in2, align=center ] 
  (in5) {$E$};
    \node[neuron, above right = 0.625cm and 1.25cm  of in5, align=center ] 
  (in6) {$F$};
  
 \draw[->, ultra thick,gray] (in1) to node[midway,below = 0.25cm]{$0.0$} (in2) ;
  \draw[->, ultra thick,blue] (in1) to node[midway,above = 0.25cm]{$0.6$}(in3);
  
  \draw[->, ultra thick,gray] (in2) to node[midway,above]{$0.0$}(in5);
  \draw[->, ultra thick,ngreen] (in2) to node[midway,left]{$0.4$} (in3);
  
  \draw[->, ultra thick,ngreen] (in3) to node[midway,above]{$0.7$}  (in4);
  
  \draw[->, ultra thick,black] (in4) to node[midway,left]{$0.7$}  (in5) ;
  
  \draw[<-, ultra thick,ngreen] (in6) to node[midway,above = 0.25cm]{$0.7$} (in4);
  \draw[<-, ultra thick,gray] (in6) to node[midway,below = 0.25cm]{$0.0$} (in5);

\end{tikzpicture}
		\caption{LU with Green-SR}
		\label{subfig:mit TE}
	\end{subfigure}

\caption{Green Segment routing example: The $A$ to $F$ traffic and the $B$ to $F$ traffic are combined to relieve the gray links.}
\label{fig: Green SR}
\end{figure}
\section{Green Segment Routing} \label{sec:GSR}
 In this section, we present our new approach for sustainable \ac{TE}. At first, we  define the green segment routing problem as a \ac{LP}. Based on this, we develop a heuristic to minimize the number of binary variables, to solve this problem.
\subsection{Green Segment Routing Problem}
We develop our approach based on the \ac{LP} by \cite{bhatia}.  It was used to keep the \ac{MLU} as low as possible. 
Figure \ref{fig: Green SR} visualizes the core idea of our approach. 
The Traffic from $A$ to $F$ is routed over $B$ and $E$ to its destination. The traffic from $B$ to $F$ is also routed over $E$, when using \ac{SPR}. When applying Green Segment Routing, the traffic from $A$ to $F$ is routed over $C$. By routing the traffic from $B$ to $F$ also over $C$, it is possible to combine both flows, and relive the links from $A$ to $B$, from $B$ to $E$ and from $E$ to $F$ from traffic.
\par 
We state the network as a directed multigraph $G=\left( V, A \right)$. The set of vertices $V$ coincides with the set of routers in the network.  As modern routers build full duplex connections, we model every connection as two directed arcs. One from router $u$ to router $v$ and vice versa. Every set of two arcs $\lbrace a_{vu},a_{uv} \rbrace$ representing one real connection, is made up of ports on the endpoints. As stated in Section \ref{subsec:power}, the number of ports related to a connection determines the capacity of an arc. Let $P\left(a\right)$ be the ports corresponding to arc $a$. The capacity provided by a port $p \in P\left(a\right)$ is denoted by $c_{p}$.  
Thus, the capacity of an arc is given by
\begin{align*}
c\left( a \right) = \sum_{p \in P\left( a \right) }  c_{p}.
\end{align*} 
Both directed arcs representing a link have the same ports and thus $c\left( a_{uv} \right) =c\left( a_{vu} \right)$. Note that while the arcs have the same capacity, their utilization can differ. There can, e.g., be far more traffic from $u$ to $v$ than from $v$ to $u$.
\par 
Every port between $u$ and $v$ belongs to two linecards, one at router $u$, one at $v$. Let $\mathcal{L}$ be the set of all linecards over all routers. Each linecard $\ell \in \mathcal{L}$ is understood as a set of ports. We introduce binary variables $\lambda_{\ell}$ and $\pi_{p}$ to denote whether a linecard $\ell$ or a port $p$ is used.  
The traffic demand from $u$ to $v$ is denoted by $t_{uv}$. To determine the amount of traffic on a specific arc, we define two precomputable functions. The first function $f_{uw} \left(a \right)$ determines the amount of traffic from $u$ to $w$ on $a$ when choosing the shortest path between $u$ and $w$. As 2-\ac{SR} is the concatenation of two shortest paths, we write 
\begin{align*}
g^{w}_{uv} \left( a \right) = f_{uw} \left(a \right) + f_{wv} \left(a \right)
\end{align*}for the amount of traffic from $u$ to  $v$ on arc $a$ when selecting intermediate router $w$. Let $x^{w}_{uv}$ denote the fraction of traffic from $u$ to $v$ that is routed over intermediate node $w$. The traffic on arc $a$ is given by 
\begin{align*}
tr \left( a \right) = \sum_{\left( u,v \right) \in V^{2}} \sum_{w \in V \backslash \lbrace u \rbrace } t_{uv} g_{uv}^{w} \left( a \right) x^w_{uv}.
\end{align*}
Further, link utilization $\text{LU}\left(a\right)$ is defined as 
\begin{align*}
\text{LU} \left( a \right) = \frac{\sum_{ \left( u,v \right) \in V^{2}}  \sum_{w \in V \backslash u} \lambda t_{uv} g_{uv}^{w} \left( a \right) x^w_{uv} }{ \sum_{p \in P \left(a\right)} \pi_{p} c_{p} }.
\end{align*}
The \ac{MLU} has to be smaller than one, to avoid congestion. Thus, we state $\theta$ as an upper bound for the \ac{MLU}. We have to choose $\theta$ in a way
 that increasing traffic can be handled, as there is some unused capacity. This comes to hand when we change to periods with more utilization. The extra capacity gives us the time to wake up all sleeping components and to restore the full capacity of the network. We will determine suitable values for $\theta$ later on. Now, we can form an ILP to estimate a linecard minimizing routing policy. The ILP is given as Problem \ref{problem:ideal_lp}. 
In Section \ref{subsec:power}, we found that the energy saving components are the linecards and ports. With $E_p$ and $E_{\ell}$ denoting the energy consumption of every port and linecard, the goal is to minimize the energy consumption of the network. The first and second constraints ensure that all traffic is routed and that no negative traffic is routed through the network. The third inequality guarantees that the link utilization for every link stays below $\theta$. The final constraint makes sure that a linecard is only turned off when every corresponding port is turned off.

\begin{problem}
\centering
\jamIfToBig{\linewidth}{
\begin{minipage}{0.85\linewidth}
    \begin{flalign*}
     & \makebox[0pt][l]{$\displaystyle{}\text{min }    \sum_{\ell \in \mathcal{L}} \lambda_{\ell} E_{\ell} + \sum_{p \in \ell} \pi_{p} E_{p}  $} \\
     & \text{s.t.} & \sum_{w \in V \backslash \lbrace u \rbrace}{x^{w}_{uv}}   &\; = \;1 & &\forall \left( u,v \right) \in V^{2}  \\ 
     & & x_{uv}^{w} &\; \geq \; 0 & &\forall  \left( u,v \right) \in V^{2}  \\
     & & \sum_{ \left( u,v \right) \in V^{2}}  \sum_{w \in  V \backslash \lbrace u \rbrace} t_{uv} g_{uv}^{w} \left( a \right) x^w_{uv}  &\; \leq \; \theta  \sum_{p \in P \left(a\right)} \pi_{p} c_{p} & &\forall a \in A \\
     & & \pi_{p} &\; \leq \; \lambda_{\ell} & &\forall \ell \in \mathcal{L} \ \forall p \in \ell \\  
     & & \pi_{p} &\; \in \lbrace 0 ,1 \rbrace \;  & &\forall p \in \bigcup_{\ell \in \mathcal{L}} \ell \\
 	 & & \lambda_{\ell} &\; \in \lbrace 0 ,1 \rbrace \; & &\forall \ell \in \mathcal{L}\\
    \end{flalign*}
\end{minipage}
}
\medskip
\caption{Green Segment Routing Problem}\label{problem:ideal_lp}
\end{problem}

\subsection{Heuristic approach}
\begin{problem}
\centering
\jamIfToBig{\linewidth}{

\begin{minipage}{0.85\linewidth}
    \begin{flalign*}
     & \makebox[0pt][l]{$\displaystyle{}\text{min }  \sum_{\pi_{p} \in \mathcal{P}} \pi_{p}  $} \\
 & \text{s.t.} & \sum_{w \in V \backslash \lbrace u \rbrace}{x^{w}_{uv}}   &\; = \;1 & &\forall \left( u,v \right) \in V^{2} \\ 
     & & x_{uv}^{w} &\; \geq \; 0 & &\forall  \left( u,v \right) \in V^{2}  \\
     & & \sum_{ \left( u,v \right) \in V^{2}}  \sum_{w \in  V \backslash \lbrace u \rbrace} t_{uv} g_{uv}^{w} \left( a \right) x^w_{uv}  &\; \leq \; \theta  \sum_{p \in P \left(a\right)} \pi_{p} c_{p} & &\forall a \in A \\
          & & \pi_{p} &\; \in \lbrace 0 ,1 \rbrace \;  & &\forall p \in \mathcal{P} 
    \end{flalign*}
\end{minipage}
}
\medskip
\caption{Port minimizing heuristic}\label{problem:LP-Ports}
\end{problem}
The ILP stated in Problem \ref{problem:ideal_lp} is rather difficult to solve, as the number of binary variables is quite high. Another problem that occurs is that the ILP assumes a fixed mapping between ports and linecards, while we assume that it is possible to reconfigure the routers to reach an energy aware configuration. This configuration has to be done only once. It would split the components of every router in two parts: One part to shut down in times of low utilization and one part to stay active throughout the whole day. 
To minimize the energy consumption, we aim to minimize only the number of active ports. This will also decrease the number of active linecards in the network, as we can turn a linecard off whenever it is possible to save enough ports on one router. This leads to Problem \ref{problem:LP-Ports}. Let $\mathcal{P}$ denote the set of all ports, over all arcs.
The first equation of this ILP defines the new goal to minimize the number of ports in the whole network. The first and second constraint ensures that the whole traffic is routed. The third inequality makes sure that the MLU stays below $\theta$. We refer to this algorithm as green 2-\ac{SR} algorithm (\emph{2SRG}). As modern hardware does not allow arbitrary splittings of traffic we state another variant of this algorithm: We may add the constraint $x_{uv}^{w} \in \lbrace 0, 1 \rbrace$. This prohibits the splitting of traffic. We refer to this variant as no splitting 2-\ac{SR} algorithm (\emph{2SRG-NS}).
\section{Data} \label{sec:Data}
In this section, we present the data used for our evaluation and the traffic analysis. We first describe the dataset provided by an  ISP. This data is used to provide an analysis of the amount of traffic over the day. This leads to a definition of what we call a low-load period. It determines the choice of the evaluation data from the ISP dataset. After that, we describe the second dataset used for our evaluation. It consists of real topologies but artificial traffic matrices. With the statistical analysis of the traffic in mind, we can modify the dataset to mimic a low-load period as well.
\subsection{ISP-Data}
The first dataset consists of real topology and traffic data collected in the backbone of an \ac{ISP}. The dataset includes snapshots of the network topology and  measured traffic matrices on a quarter-hour resolution measured with the method stated in \cite{methode-horneffer}. The topology of the network changes over time as the network grows continuously. To define the low-load period, we used all snapshots available to us from 2020. As a result, our traffic analysis is based on over 35000 quarter-hour snapshots. As mentioned before, the network has grown over the year, especially in the second half of the year. Thus, the number of nodes varies from about 160 to 190 active nodes and the number of active links has grown from about 3700 active ports to 4600 ports.
\subsection{Repetita Data}
Our second dataset features ten selected instances from the publicly available Repetita \cite{repetita} dataset (c.f. Table \ref{tab:repetita_instances}). All the data used in this work can also be found in the topology zoo \cite{topo-zoo}. It contains several real-world topologies, with artificial traffic matrices. The goal of the project is to create a public available dataset to test and optimize traffic engineering approaches.
\par 
These instances are very challenging to optimize in terms of the \ac{MLU}. The artificial traffic matrices are tailored to result in a MLU of 90 \% with optimal routing. Thus, there is no room for power saving approaches.  We decided to scale down the traffic matrices. We choose the scaling factor $0.5$ to create instances that capture the effect of low utilization. 
\par
This scaling factor reflects the behavior of the ISP data during phases of low utilization. We will examine this further in Section \ref{sec:lowpeak}.
\par 
The topologies do not contain information about parallel links. Thus, we had to make some assumptions regarding the number of ports and number of linecards used in these topologies. We assumed that every link in the Repetita data consists of $4$ parallel links, each referring to one port. This is the average number of parallel links in the ISP dataset. For every $8$ ports per node we added one linecard. This assumption coincides with the example router configured in section \ref{subsec:power}. With these changes on the dataset, we mimic the same behavior as the real-world dataset and evaluate our approach on different topologies.  
\begin{table}
	\centering
	\caption{Overview over the selected Repetita instances.}
	\label{tab:repetita_instances}
	\scriptsize
	\begin{tabular}{l c c c}
		\toprule
		Name & Nodes & Edges & Identifier\\
		\midrule
		DeutscheTelekom & 30 & 110 & A\\
		Forthnet & 62 & 124 & B\\
		Globenet & 67 & 226 & C\\
		GtsCzechRepublic & 32 & 66 & D\\
		RedBestel & 84 & 202 & E\\
		Renater2008 & 33 & 86 & F\\
		Renater2010 & 43 & 112 & G\\
		Ulaknet & 82 & 164 & H\\
		Uninett2010 & 74 & 202 & I\\
		Uunet & 49 & 168 & J\\
		\bottomrule
	\end{tabular}
\end{table}

\section{Definition of the Low-Load Period} \label{sec:lowpeak}
\begin{figure}

\includegraphics[width=0.9\linewidth]{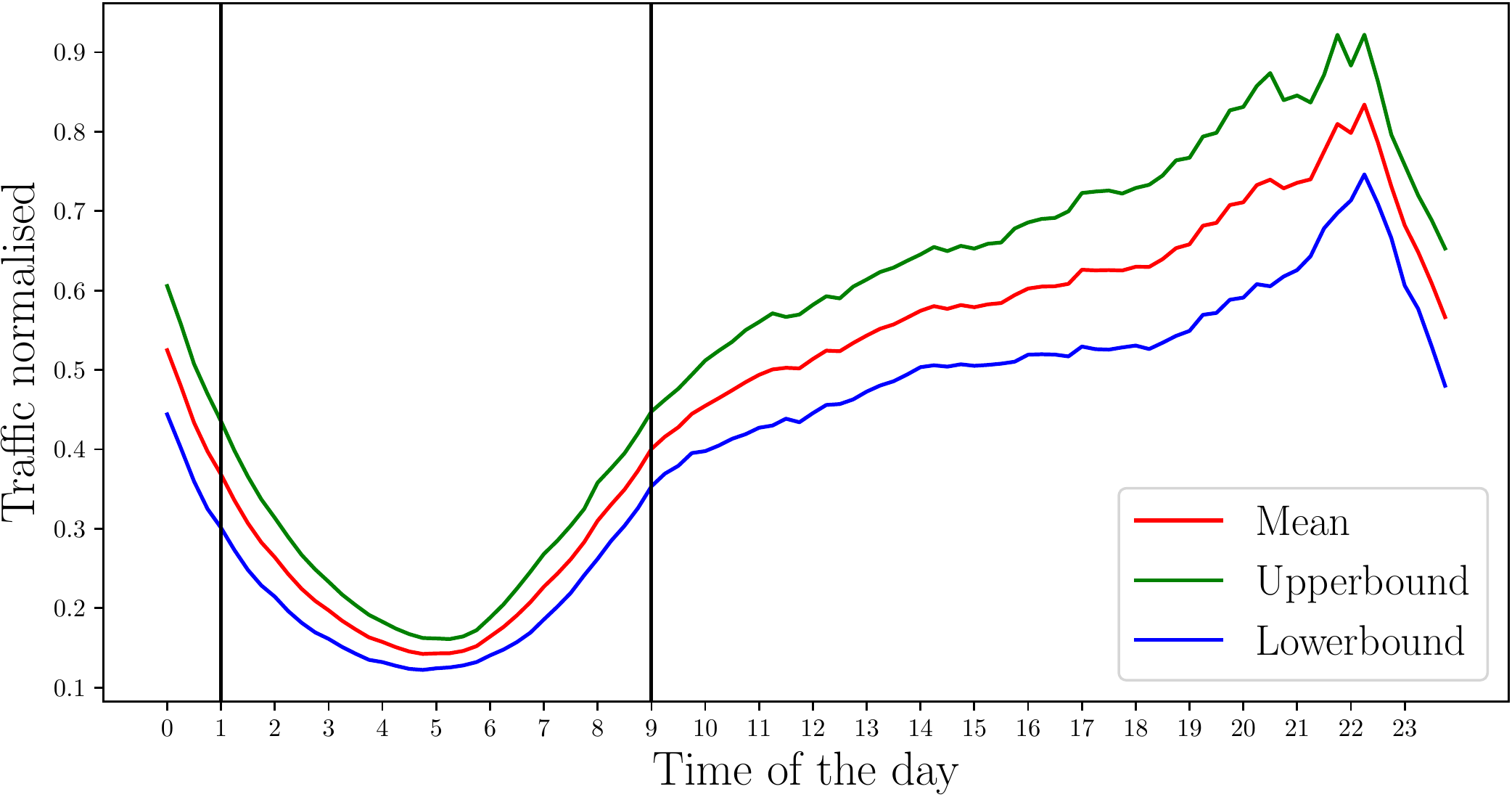}
\caption{Normalized Traffic 2020 with Low-Load Period}
\label{fig:traffic2020}
\end{figure}
In this section, we  provide a detailed analysis of the daily traffic volume. This will help us to select appropriate scenarios for evaluation. The traffic volume should be low enough to have a saving potential. Further, it should be long enough to spare reconfiguration effort. We choose to model the traffic as a gaussian distribution. The traffic to a given time point $t$ is the realization of a gaussian distributed random variable. The deviation $\sigma \left( t \right) $ and the expected value $\mu \left( t\right)$  are time-dependent functions
\begin{align*}
Tr \left( t \right)  \sim \mathcal{N} \left( \mu \left( t \right), \sigma \left( t \right)  \right).
\end{align*}
To estimate the deviation and the expected value, we used the corresponding empirical estimators 
\begin{align*}
\mu \left( t \right) = \frac{1}{n} \sum_{i=1}^{n} Tr \left( t \right)
\end{align*}
and
\begin{align*}
\sigma \left( t \right) = \sqrt{ \frac{1}{n} \sum_{i=1}^{n} \left( Tr \left( t \right) - \mu \left( t \right) \right)^{2}}.
\end{align*} 
All snapshots from 2020 were used to calculate both values. With the expected value and the deviation, it is possible to estimate the probability of a certain amount of traffic. We calculated a $0.7$ confidence interval of the traffic volume. The results can be found in Figure \ref{fig:traffic2020}. The red line marks expected traffic. The blue line marks the lower bound and the green line marks the upper bound of the confidence interval. 
\par
 The amount of traffic is below $50$ \%  of the daily maximum from 1:00 to 9:00. We state this interval as the low-load period. In this period, the amount of traffic is low enough to have a sufficient saving potential. Further, the period is long enough to spare reconfiguration effort. 
To evaluate the differences within the period, we decided to take a snapshot from the beginning or the end and one from the middle. We state the time from 1:00 to 1:30 as the beginning, and the time from 8:00 to 9:00 as the end of the period. The middle of the period is the time from 4:00 to 5:45. 
\par
With the low-load interval defined, we now need to choose an upper bound for the \ac{MLU}. The \ac{MLU} is a linear function. Thus, if the total traffic changes by a factor of $\lambda$, the load factor also changes by a factor of $\lambda$. 
With this in mind, it is possible to set an upper bound for the \ac{MLU}.
\par
As we stated before, the amount of all traffic is with probability of $85$ \%  below the upper bound. Thus, we need some extra capacity to cope with these spikes. We decided to choose an upper bound of $70$\%  for the \ac{MLU}. This allows an increase of $40$\%  for all traffic demands before congestion occurs.
A higher \ac{MLU} decreases the ability to cope with spikes on an enormous level. When allowing an \ac{MLU} up to $90$\%  we can deal with spikes up to $10$\%, if we allow $80$\%  we can handle spikes up to $20$\%. A \ac{MLU} of $70$\%  is a good compromise between avoiding congestion and still having enough capacity left to turn unused components off. 

\begin{figure*}[]
	\centering
	\begin{subfigure}{0.31\linewidth}
		\centering
		\includegraphics[width=.98\linewidth]{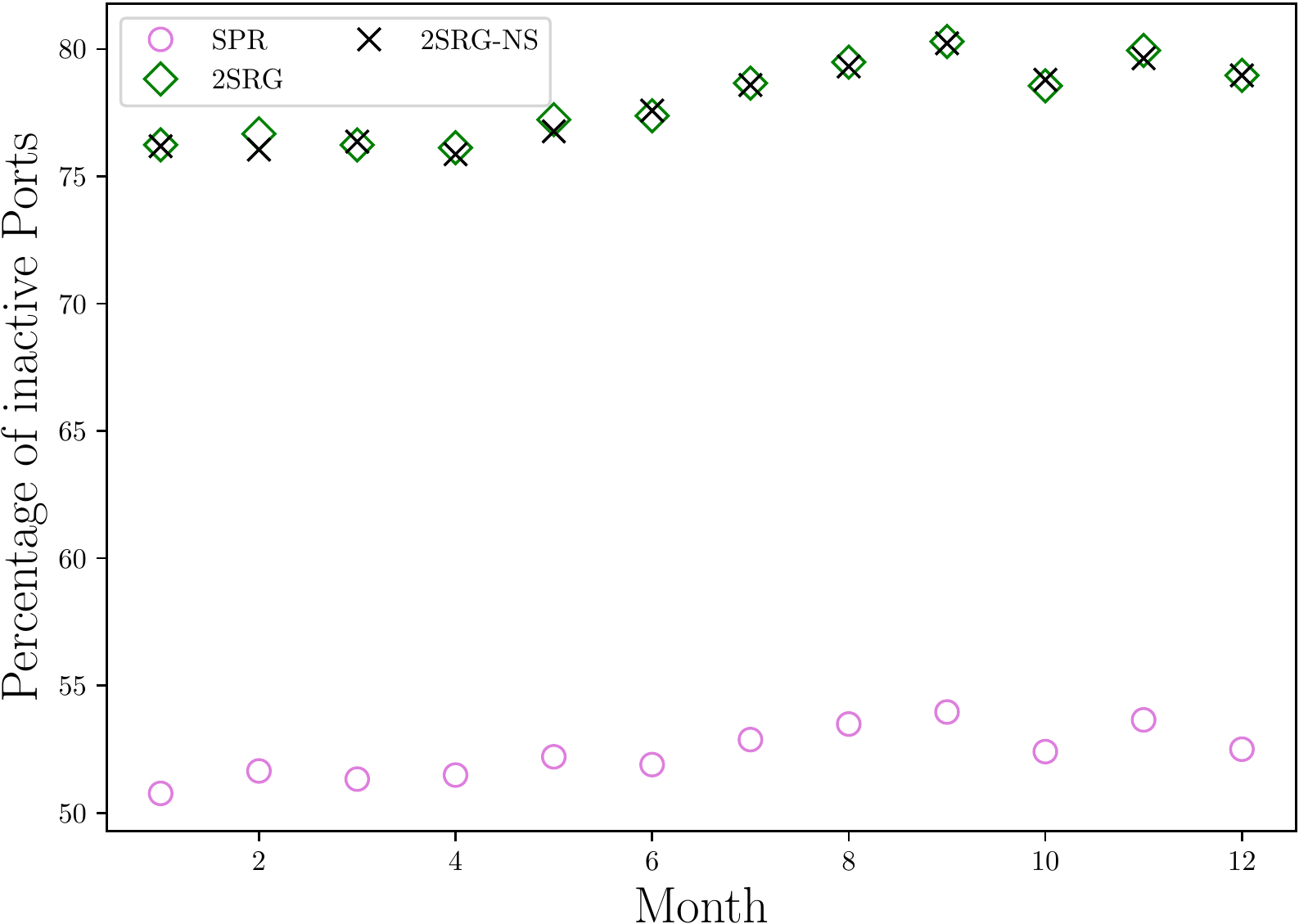}
		\caption{ISP-Data from 4:00-5:45}
		\label{subfig:ports-mitte}
	\end{subfigure}
	\begin{subfigure}{0.31\linewidth}
		\centering
		
		\includegraphics[width=.98\linewidth]{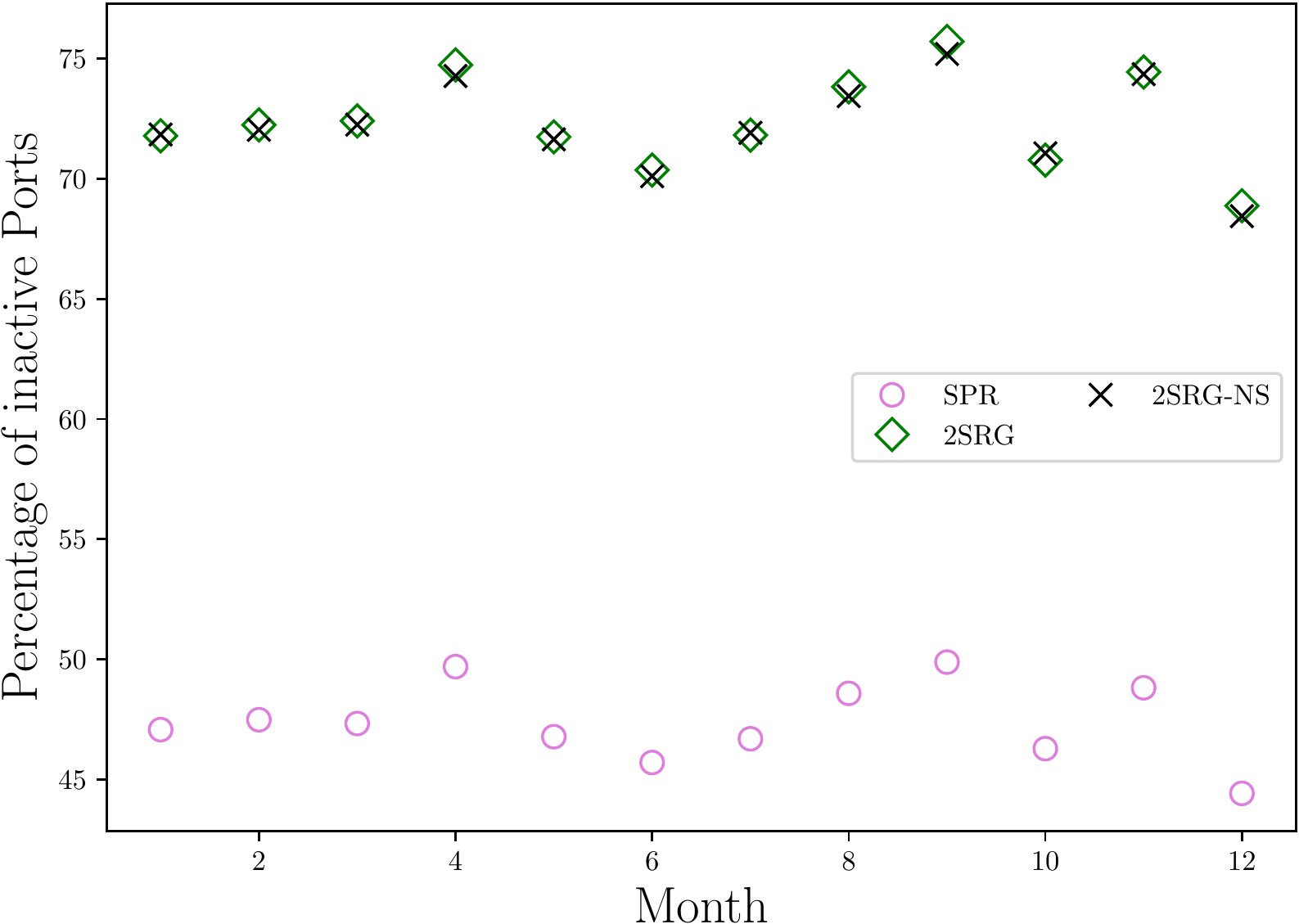}
		\caption{ISP-Data from 8:00-9:00 and 00:30-1:00}
		\label{subfig:ports-rand}
	\end{subfigure}
	\begin{subfigure}{0.31\linewidth}
		\centering
		\includegraphics[width=1.0\linewidth]{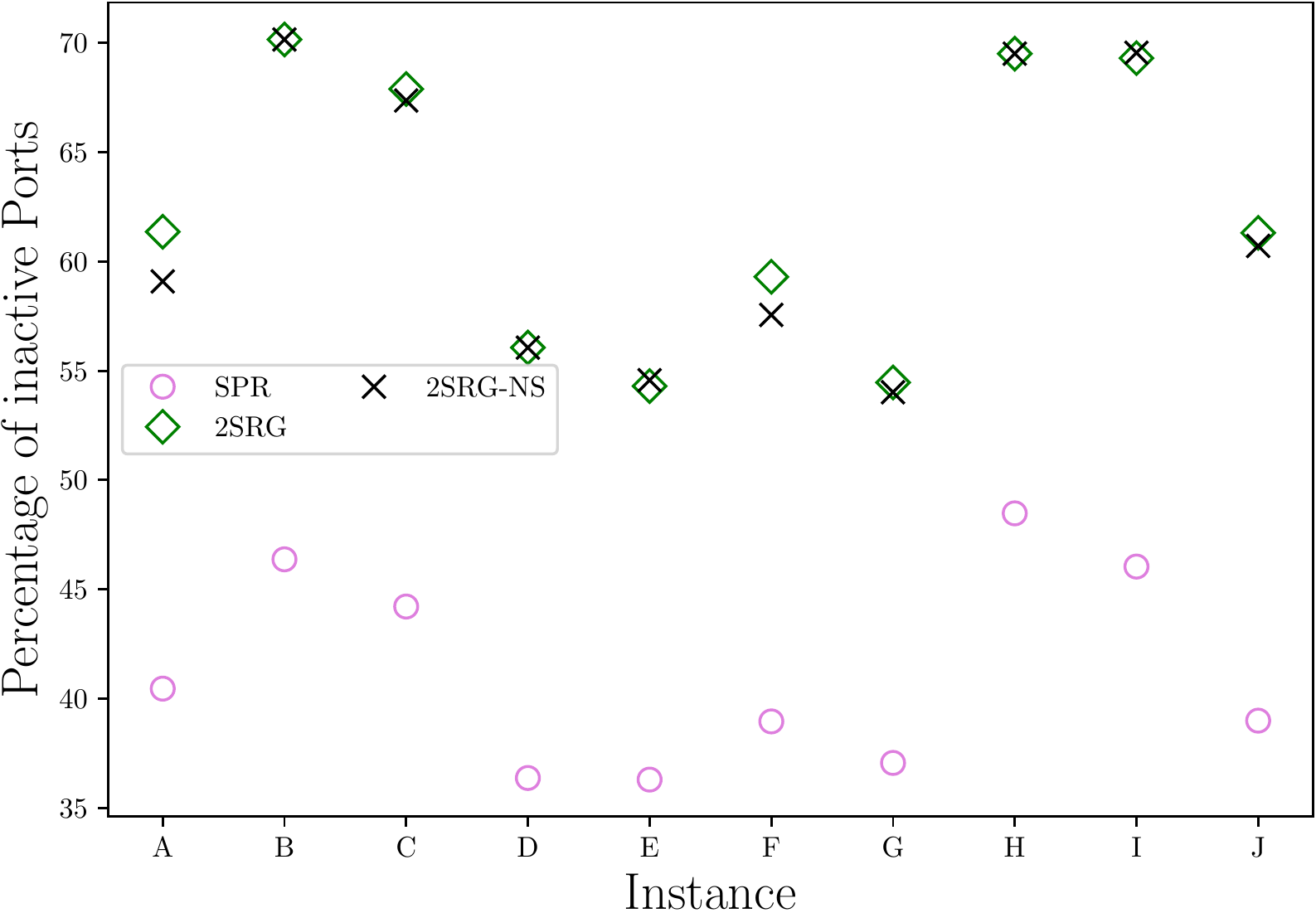}
		\caption{Repetita}
		\label{subfig:ports-repetita}
	\end{subfigure}
	\caption{Percentage of inactive Ports }
	\label{fig:Ports}
\end{figure*}
\section{Evaluation } \label{sec:eval}

In this section, we evaluate the performance of our newly
developed traffic engineering approach regarding the achievable number of inactive linecards and ports. We also take a look at the \ac{MLU} after turning ports and linecards off.
\subsection{Algorithms}
To evaluate the quality of our approach, we compared our algorithms, \emph{2SRG} and \emph{2SRG-NS}, with \ac{SPR}. It represents the behavior of methods that are based on the calculation of shortest paths, such as \ac{OSPF} and \ac{IS-IS}. With our evaluation of  of \ac{SPR} we gain insights into the actual state that we want to improve with our new approach.
\par
To evaluate the energy saving potential of this algorithm, we calculate a \ac{SPR} solution. Afterwards, we take a look at the utilization of every link. If the link utilization is below the threshold $\theta$, there is unused capacity on the link. If the unused capacity, corresponds to the capacity of one or more ports, we put the corresponding number of ports in an inactive state.  For every set of eight unused ports on the same router, we assume that we can switch one linecard off. 
\par 
In our analysis of the traffic throughout one day, we showed that for eight hours a day, the possibility that the traffic exceeds $50$\%  of the daily maximum is below $85$\%. To make sure that a bit of extra traffic can be handled and that there is some capacity left for possible link failures, we set $0.7$ as the upper bound for the link utilization, following the same  line of arguments stated in section \ref{sec:lowpeak}. 
\par
To solve the ILP stated in Problem \ref{problem:LP-Ports} we need to apply an heuristic approach. It was not possible to find an optimal solution for the \ac{ISP} instances, even with a few days computation time. Thus, we decided to use a rounding approach to calculate a feasible solution. We treated the number of ports as a continuous variable. However, we now need to post-process the optimization results, from a continuous solution back to an integer one that can actually be deployed in practice. Thus, we modified the solution via rounding to the next possible integer solution. This is an effective way to obtain a feasible  solution.

\subsection{Port-saving potential} \label{subsec:port potential}

\begin{figure*}
	\centering
	\begin{subfigure}{0.31\linewidth}
		\centering
		\includegraphics[width=.98\linewidth]{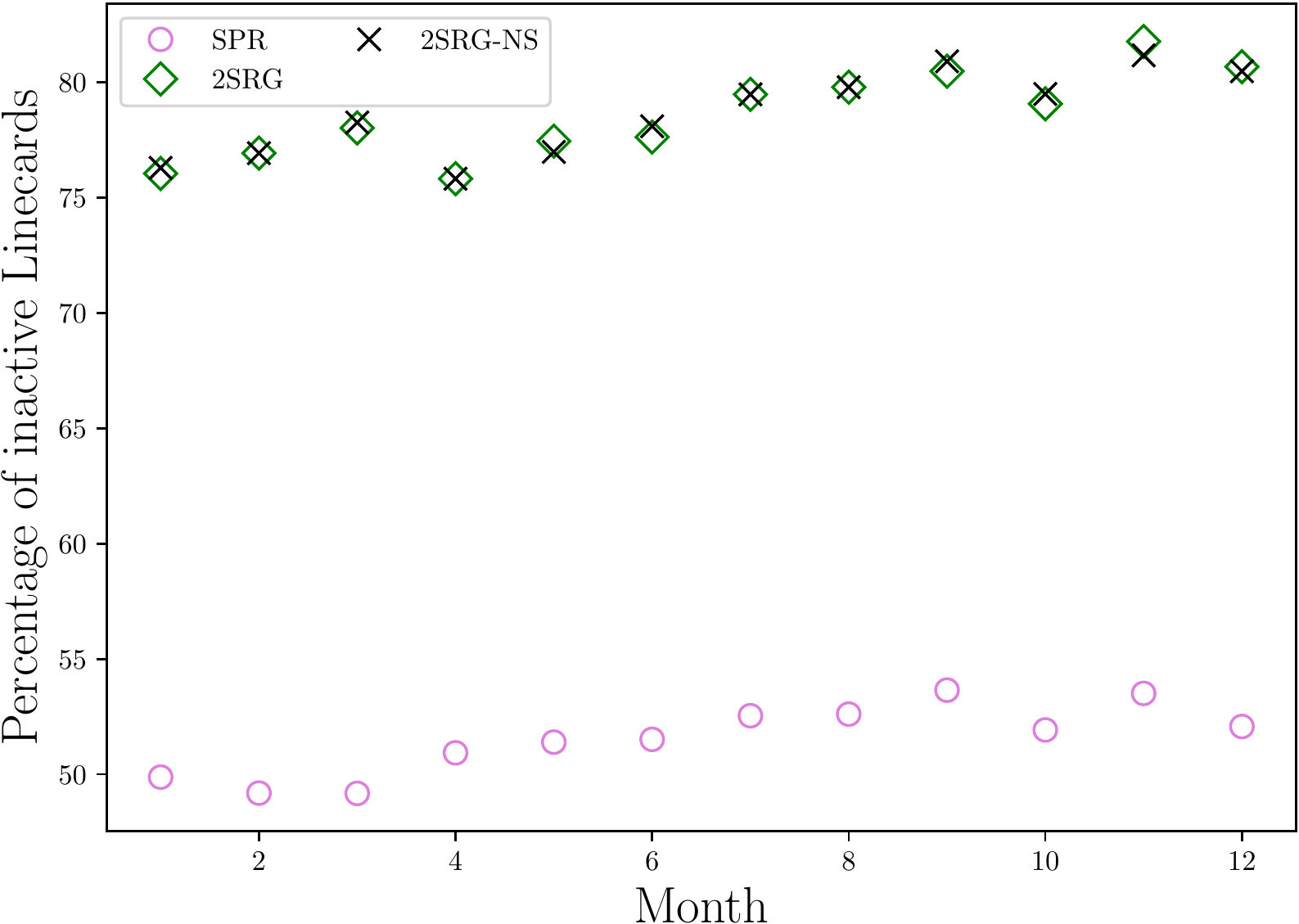}

		\caption{ISP-Data from 4:00-5:45}
		\label{subfig:lc-Mitte}
	\end{subfigure}
	\begin{subfigure}{0.31\linewidth}
		\centering
		\includegraphics[width=.98\linewidth]{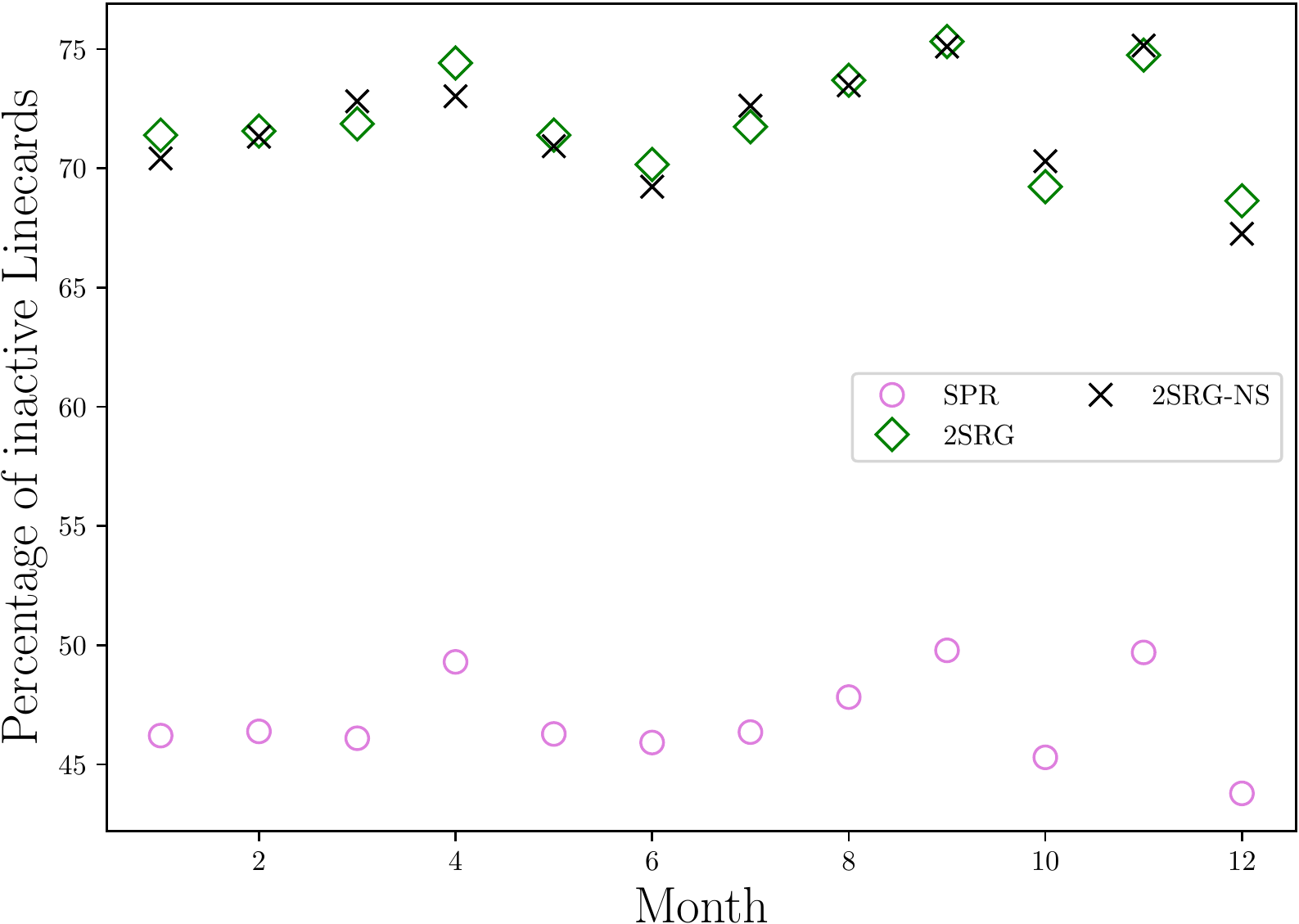}
		\caption{ISP-Data from 8:00-9:00 and 00:30-1:00}
		\label{subfig:lc-Rand}
	\end{subfigure}
	\begin{subfigure}{0.31\linewidth}
		\centering
		\includegraphics[width=1.0\linewidth]{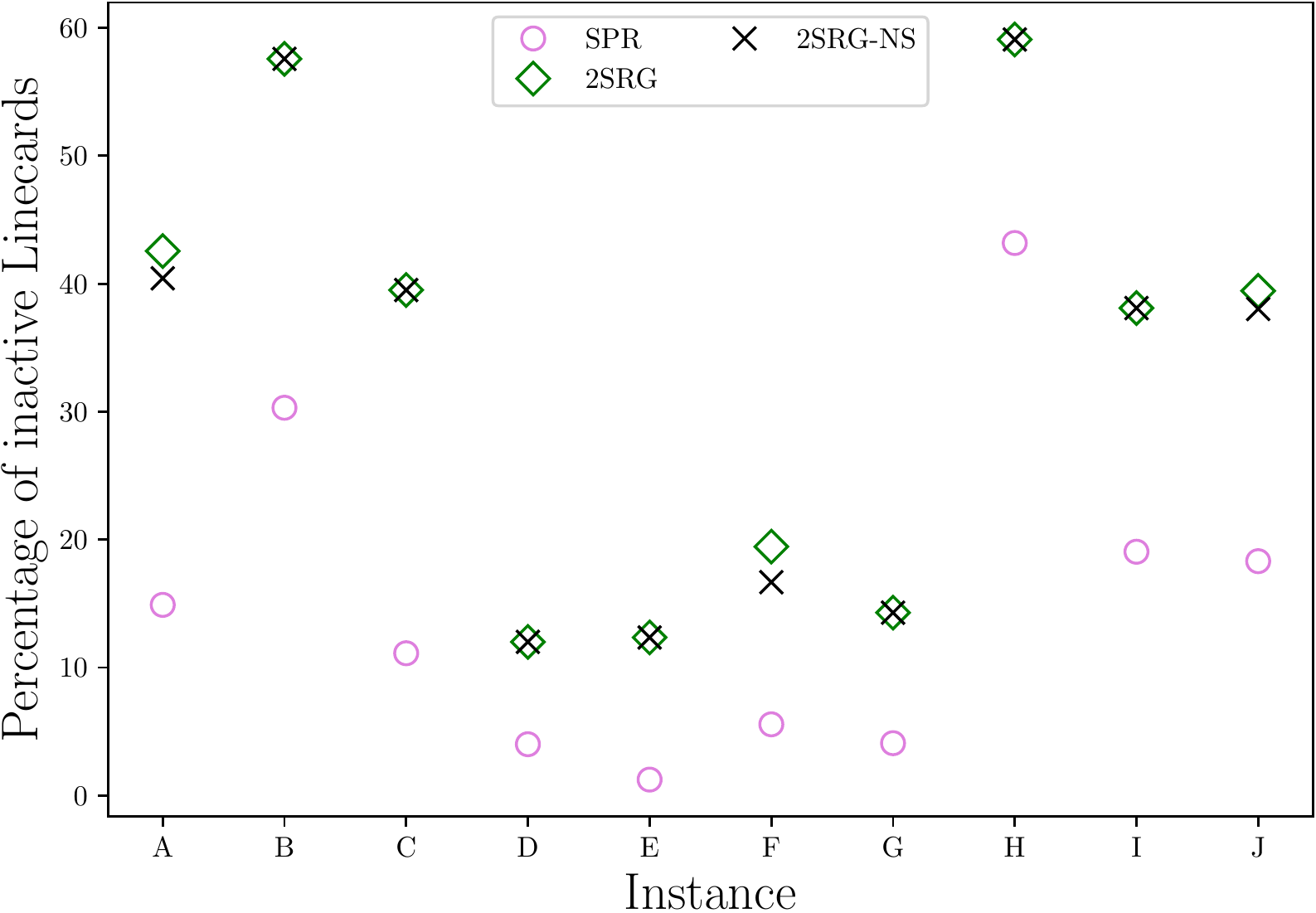}
		\caption{Repetita}
		\label{subfig:lc-repetita}
	\end{subfigure}
	\caption{Percentage of inactive linecards}
	\label{fig:LC}
\end{figure*}
To assess the port saving capabilities of our new approaches, we optimized each instance from our reference datasets with
it. The results can be found in Figure \ref{fig:Ports}.
\par
In the time between 4:00 and 05:45  the number of unused ports varies from 75\% to 80\%, (c.f. \ref{subfig:ports-mitte}). It can be observed that there is nearly no difference between the no splitting approach and the arbitrary splitting approach. This is explainable via the rounding heuristic used to solve the \ac{LP}. We treated the number of ports as a continuous variable. The continuous solution had to be rounded down to an integer solution. Thus, the rounding loss of \emph{2SRG-NS} was smaller than the rounding loss of \emph{2SRG}. This was expectable  because the solution space of \emph{2SRG} is larger. In most cases, however, this did not have an effect. This shows that even with one policy per demand, our approach can eliminate a significant proportion of ports.
When looking at SPR, around 50\% of all ports could be turned off in most instances. 
\par
The results for the times that belong to the edge of the low-load period can be found in Figure \ref{subfig:ports-rand}. Our new approaches manage to set $70$\%  of all ports into an inactive state. \ac{SPR}  manages to set $45$\% of all ports to an inactive state.  
We have shown that there is a massive amount of unneeded capacity in the network in times of low traffic demands, and that it is possible to find a routing policy which is capable to take advantage of that. Furthermore, we can identify the corresponding ports, causing the amount of unused capacity. 
\par
We were able to show that more than 70 \% of all ports can be turned off for both time periods. Despite the increase in data traffic, the proportion of unused ports has only decreased by $5$\%. As all traffic could be transported when using the solution from the edge of the low-load period for the entire period, it is possible to use this solution eight hours a day.
\par
The number of unused ports within the Repetita datasets differ a lot more, see Figure \ref{subfig:ports-repetita}.
For instances B, H, and I \emph{2SRG} and \emph{2SRG-NS} managed to set $70$\%  of all ports in an inactive state. For the rest of the instances, we do not need at least $50$\%  of all ports. The new algorithms manage to set additional $20$\%  of all ports in an inactive state compared to and \ac{SPR}. The number of inactive ports differs that much, because the topologies are different. For example, the Telekom, Forthnet, and Globenet instances are very dense. This means that the number of connections is much higher in relation to the number of all possible connections. Thus, we can choose between different paths. Other instances like the GTSCzechRepublic net are not that dense. Therefore, the algorithm has fewer paths to choose from, and thus there is less potential for optimization.
\par
Overall, we observed that the \emph{2SRG} and \emph{2SRG-NS} can set more ports to an inactive state than \ac{SPR}. This result is unsurprising, as \ac{SPR} does not minimize towards this goal. Nevertheless, the evaluation of SPR provides an insight into the standard currently in use.
\subsection{Linecard-saving potential} \label{subsec:Lc}
\begin{figure*}
	\centering
	\begin{subfigure}{0.31\linewidth}
		\centering
		\includegraphics[width=.98\linewidth]{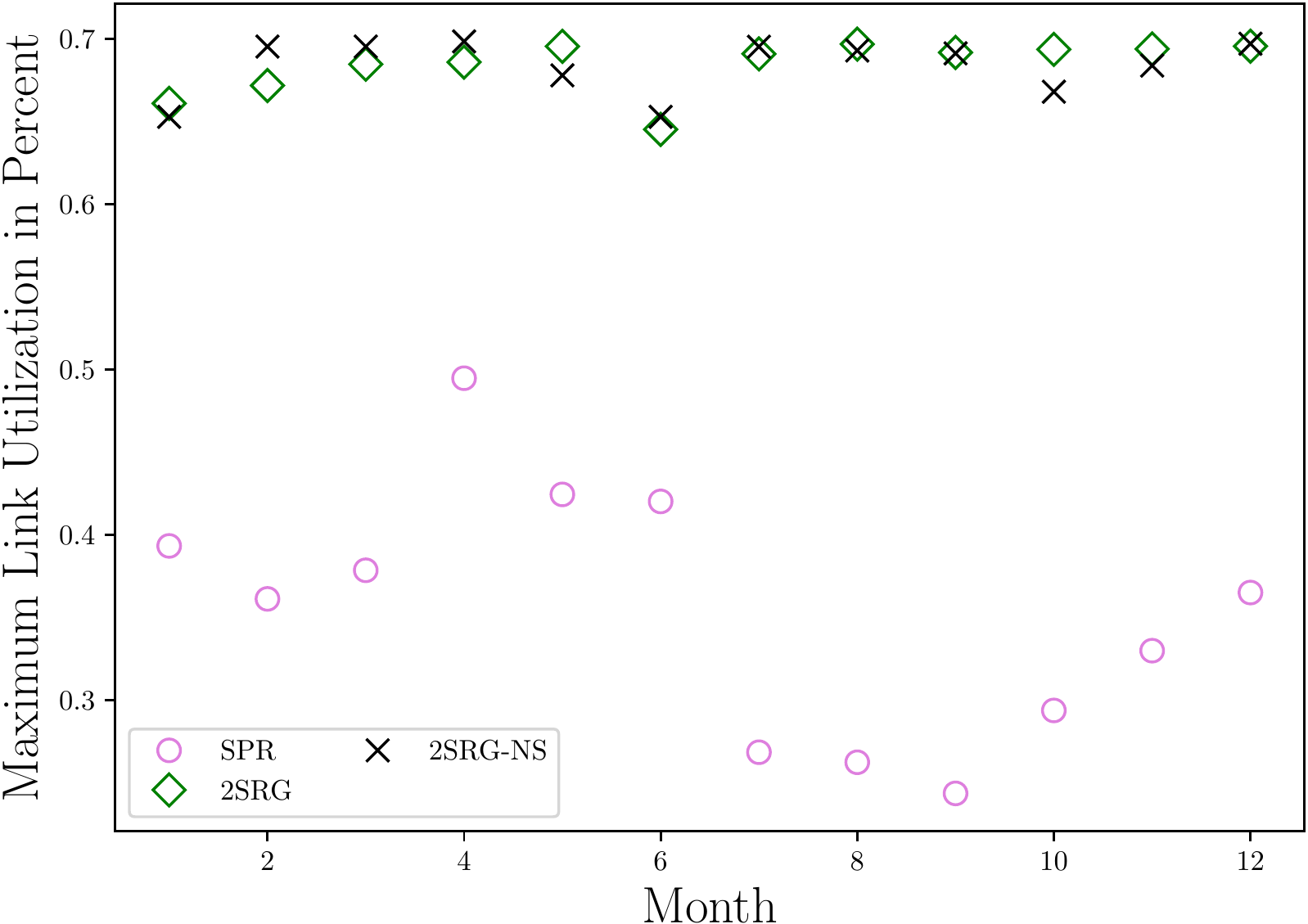}
		
		\caption{ISP-Data from 4:00 and 05:45}
		\label{subfig:MLU-Mitte}
	\end{subfigure}
	\begin{subfigure}{0.31\linewidth}
		\centering
		\includegraphics[width=.98\linewidth]{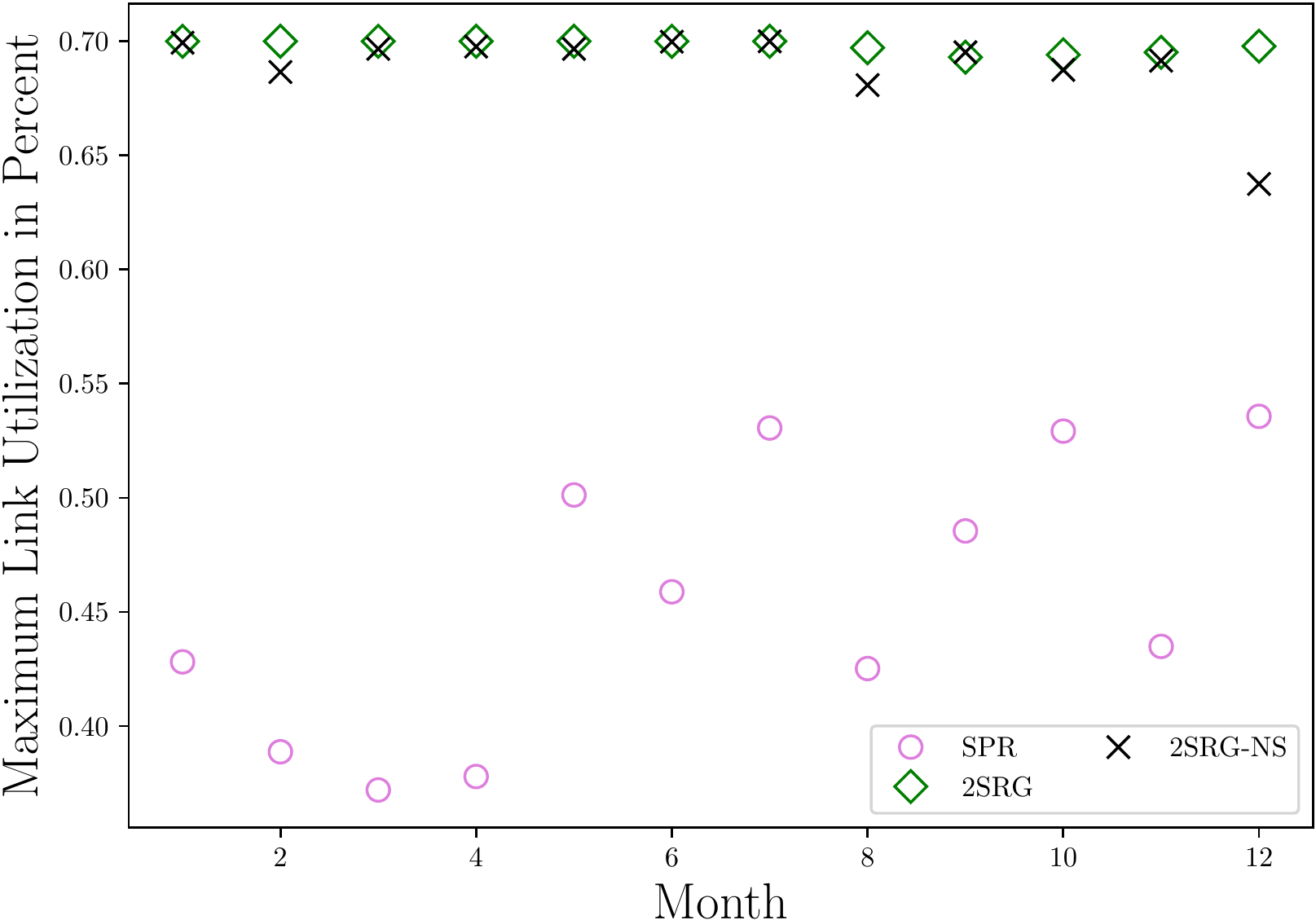}
		\caption{ISP-Data from 8:00-9:00 and 00:30-1:00}
		\label{subfig:MLU-Rand}
	\end{subfigure}
	\begin{subfigure}{0.31\linewidth}
		\centering
		\includegraphics[width=1.0\linewidth]{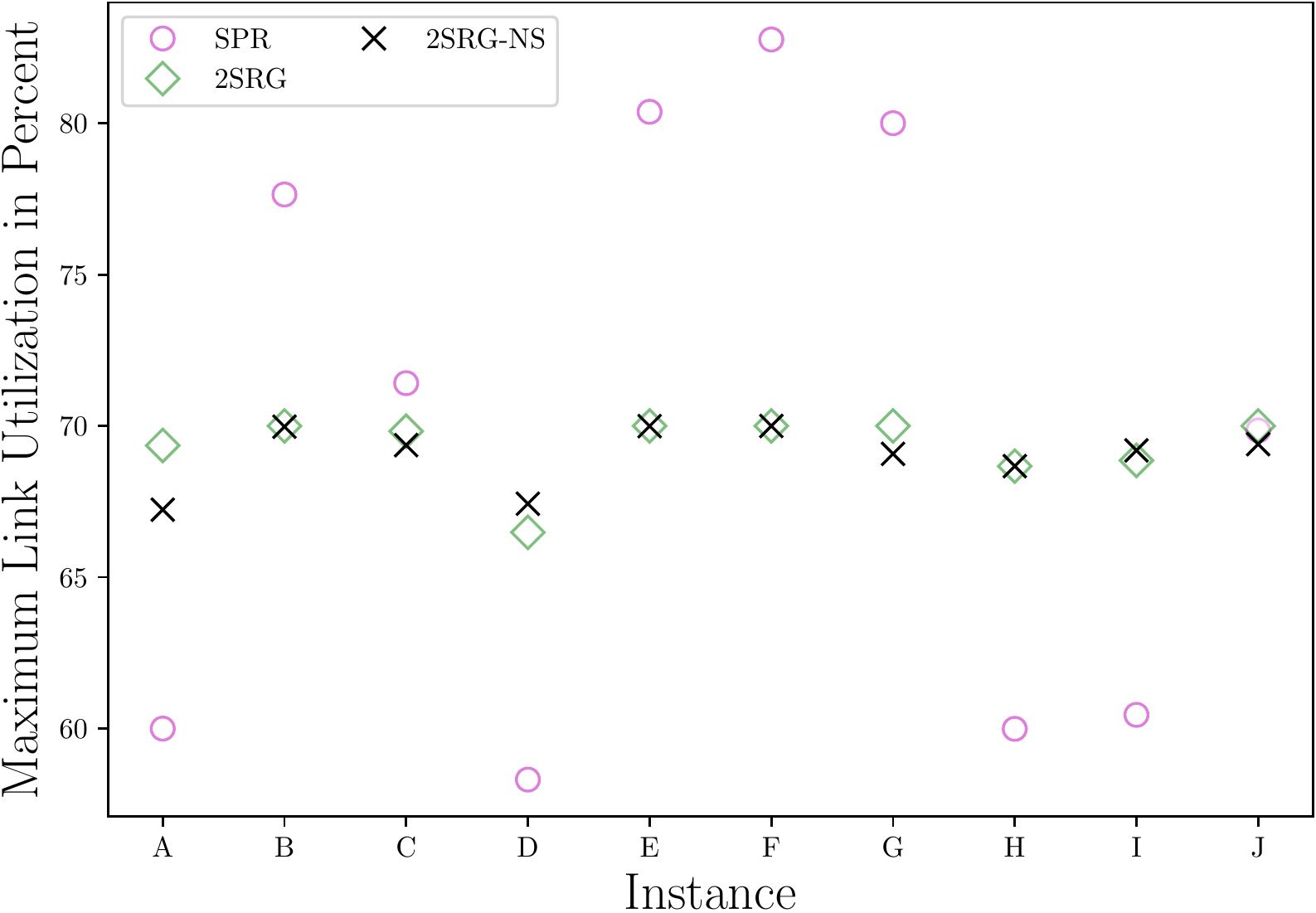}
		\caption{Repetita}
		\label{subfig:MLU-Repetita}
	\end{subfigure}
	\caption{MLU after setting ports in an inactive state}
	\label{fig:MLU}
\end{figure*}
In Section \ref{subsec:power}, we stated that the linecards consume the majority of the energy within the network. Therefore, we evaluated the number of inactive linecards depending on the routing algorithm. We assumed that for every eight ports per router, we can turn one linecard off. In practice, this makes it necessary to change the port linecard mapping. But this has to be done only once, if a routing policy for the low-load period is found.
Furthermore, we assume that only linecards with eight ports are used in the network. We are only counting the numbers of linecards that provide services within the network. For example, when there are $14$ ports used for connections within the backbone and six ports used to connect an access network, we assume that we can at most set one linecard in an inactive state, even though there are clearly three linecards in use. Two linecards are needed to keep the network connected.
\par
In Figure \ref{subfig:lc-repetita}, the percentage of unused linecards within the topologies provided by the Repetita dataset can be found. The results of \emph{2SRG} and \emph{2SRG-NS} differ strongly based on the different topologies. The results varies from  $10$\% to $70$\% unused line cards. 
The result of \ac{SPR} also varys depending on the topology. In particular, it is notable that few unused linecards can be found for instances D, E, F, and G. The number of unused ports was also lower for these topologies compared to the other topologies. The effect was even stronger regarding the number of inactive linecards, since often not enough ports per router were inactive to set a linecard inactive. 
\par
The results regarding the number of unused linecards in the middle of the low-load period can be found in \ref{subfig:lc-Mitte}. The proportion of inactive linecards is about the same as the proportion of inactive ports.  Thus, the \emph{2SRG} and \emph{2SRG-NS} algorithms manage to set more linecards in an inactive state than \ac{SPR}. Both algorithms, \emph{2SRG} and \emph{2SRG-NS}, manage to set between $75$\%  and $80$\%  of the linecards in an inactive state. The \ac{SPR} algorithm manages to turn around $50$\%  of the linecards off.  
\par 
The results obtained by optimizing the \ac{ISP} data from the time between 00:30h to 1:00h and 8:00h to 9:00h, reflect the port result as well, see Figure \ref{subfig:lc-Rand}. Both of our new approaches manage to shut down around $70$\%  of all linecards. There is still nearly no difference between the \emph{2SRG-NS} and the \emph{2SRG} approach. It is worth noting that traffic has doubled compared to the absolute minimum, and we can still shut down $70$\%  of all linecards. \par 
The differences between the edge of the period and the center are only slight. This implies that a solution for the period of eight hours can be used, as a finer subdivision of the interval does not promise a large additional gain.
\par
Note that the fraction of inactive linecards is nearly as high as the fraction of inactive ports for all \ac{ISP} instances. This indicates that by minimizing the fraction of active ports in the ISP network, we reduce the quota of inactive linecards
in the same manner. We have shown that the idea of minimizing the number of linecards indirectly by minimizing the number of ports within the network is a suitable approach to make the problem easier to solve.
\par
As we stated in \ref{subsec:power} the linecards use up to $80$\%  of the energy within a router and, hence, the linecards are responsible for $80$\%  of the energy consumption within the network. By turning off $70$\%  of all linecards we can save up to $56$\%  of the energy used within the network without turning a router off. This shows that our algorithm can save a huge amount of energy under the assumption, that a linecard can be set to an inactive state. Yet, this does not hold true for all linecards. But as modern linecards are built with a power-saving mode, it is only a matter of time before these assumptions apply to more networks.
\subsection{Maximum Link Utilization} \label{MLU}
In this section, we take a look at the resulting MLU of every approach. The MLU is crucial as the network needs to be capable to deal with sudden traffic spikes. We calculated the \ac{MLU} after setting the unneeded ports in an inactive state.  
The  MLU for all instances is at a constant level of about $70$\%  for \emph{2SRG} and \emph{2SRG-NS}. This is because the algorithm can optimize the number of inactive ports, as long as the utilization is under $70$\%. Therefore, the optimizer always uses the permitted \ac{MLU} to have fewer active ports in the network.
\ac{SPR} did not manage for  all Repetita instances to keep the utilization under $70$\%. This is explainable, as the traffic on the instances for Repetita is designed to lead to be challenging in terms of the MLU. It is not unusual that the \ac{MLU} with non modified traffic exceeds $200$\%. Nevertheless, it was possible with \ac{SPR} to save some ports as we are only looking at the \ac{MLU} of the entire network. Individual links might have a utilization of less than $70$\% and, hence, some unneeded capacity that can be switched off.
\par
With our approach, it was possible to find a solution for all instances that was the best regarding inactive components. Furthermore, we could provide a solution with a lower MLU than SPR for some instances. Thus, we were able to demonstrate that our approach not only requires fewer components to route the same traffic, but also achieves a lower utilization of the components in some cases. 
\par
Overall, the \emph{2SRG-NS} algorithm performs as good as \emph{2SRG}, even though it cannot split the traffic in an arbitrary manner. This holds for every aspect of our evaluation. So prohibiting  splitting has no negative effect on the MLU or the number of inactive components. This shows that our approach can be limited to one policy per demand. Which makes it a lot easier to implement in existing hardware.

\subsection{Computation Times and Resource Demands}
The LP-based approaches are quite demanding in terms of computation time and resources. For the ISP instances and the larger Repetita instances, the calculation of the \emph{2SRG-NS} took several hours. They also needed several hundred Gigabyte of RAM. As we aim to optimize the network on a long-term, this is no problem. All computations are done on a computer with two AMD EPYC 7452 CPUs, 256GB of RAM and 64-bit Ubuntu 20.04.1.
The LPs are solved using CPLEX \cite{cplex}.  

\section{Conclusion and Future Work}
In this paper, we discussed the usability of \ac{SR} for green \ac{TE}. We developed a \ac{TE} approach, that on the one hand allows to identify unused components and on the other hand can keep the link utilization under a fixed level. Based on a statistical analysis of the amount of traffic, we developed an eight-hour daily interval in which the amount of traffic remains below $50$\%  of the daily maximum with a certainty of 85\%. This time interval is sufficient to make the effort of configuration worthwhile, and the traffic is low enough to switch off hardware components.
After first formulating an optimal LP, two further LPs were developed based on this LP to minimize the energy consumption of a backbone network. We chose to tackle the goal of turning off linecards, since the linecards of the routers are responsible for $80$\%  of the power consumed by a network. We did so, by turning off as many ports as possible within a network to reduce the number of binaries in our \ac{LP}. 
We showed that this heuristic can set nearly the same fraction of linecards in an inactive state than the fraction of ports. This implies that this is a suitable way to minimize the number of binary variables in the ILP. We have shown that in a real-world scenario, we could possibly save up to $56$\%  of the energy consumed by the network, under certain hardware assumptions. Moreover, with our approach even when the traffic has doubled, the number of active increased by $5$\%. This strongly suggests that a solution can be used for the entire low-load period. In future work, we have to take a closer look into this aspect. More precisely, it is important to make the reconfiguration intervals as long as possible, while still keeping the proportion of saved energy as high as possible.
We were able to confirm these results on other topologies with artificial traffic matrices. Our approach could achieve a lower MLU and required fewer active components than SPR in some cases.
In future work, additional real-world constraints must be integrated to implement our approach in real-world networks.  For example, one aspect that we have omitted in this work is a delay constraint. We are planning to tackle these additional constraints in the future. 
Overall, our results show that \ac{SR} can be used for green \ac{TE}. They give a good orientation what can be achieved with \ac{SR} in this field.

\section{Acknowledgment}

This work was supported in part by the German Research Foundation (DFG), Project No. AS 341/7-1.


\bibliographystyle{IEEEtranS}
\bibliography{IEEEabrv,quellen}
\balance
\end{document}